# Nonthermal emission in the central starburst region of M82

M. Persic,[1] R. Rando[2] and Y. Rephaeli[3]

[1] Istituto Nazionale di Astrofisica, Padova Astronomical Observatory, vicolo dell'Osservatorio 5, I-35122 Padova, Italy
Istituto Nazionale di Fisica Nucleare, sezione di Trieste (Gruppo collegato di Udine), I-34127 Trieste, Italy
e-mail: massimo.persic@inaf.it

[2] Department of Physics & Astronomy, University of Padova, via F. Marzolo 8, 35131 Padova, Italy
Istituto Nazionale di Fisica Nucleare, sezione di Padova, I-35131 Padova, Italy
Center of Studies and Activities for Space, University of Padova, via Venezia 15, I-35131 Padova, Italy
e-mail: riccardo.rando@unipd.it

[3] School of Physics & Astronomy, Tel Aviv University, Tel Aviv 69978, Israel
Department of Physics, University of California at San Diego, La Jolla, CA 92093, USA
e-mail: yoelr@tauex.tau.ac.il

**ABSTRACT**

*Context.* Diffuse nonthermal (NT) emission from the central starburst (CSB) region of M82 has been measured at radio, X-ray, and $\gamma$-ray energies. Far-infrared (FIR), radio, and X-ray emission maps are mutually consistent, with the radio and X-ray emissions being spectrally similar. These observational results suggest that NT X-ray emission is likely produced in Compton scattering of radio emitting electrons off the ambient FIR field. We present results of our analysis of 16.3 years of *Fermi*-LAT measurements, which – combined with the newly published, improved VERITAS point-source data – constitute the deepest, most extensive currently available $\gamma$-ray dataset on M82.

*Aims.* We aim to self-consistently model the NT radio to $\gamma$-ray spectral energy distribution (SED) of the CSB as emission by relativistic electrons and protons. Key features of our models are the use – for the first time in a broadband NT spectral study of a starburst galaxy – of diffuse X-ray and radio emission from the CSB region, which allow for an overall calibration of the electron spectrum, and the identification of the $\gtrsim 50$ GeV emission as pionic in origin. This enables the determination of the zero-point and slope of the proton (and secondary electron) spectrum, and meaningful estimates of the energy densities of particles and magnetic fields.

*Methods.* We consider all relevant radiative and adiabatic processes involving relativistic and thermal electrons and protons. We use detailed descriptions of the radiation fields in the CSB region, the most important of which is the local FIR field, a graybody parametrized by the dust emission index and temperature ($\beta$, $T_{\rm d}$).

*Results.* Our SED modeling indicates that *(i)* the $\gtrsim 10$ GeV emission is mostly pionic, *(ii)* the $0.1 \lesssim E_\gamma/{\rm GeV} \lesssim 10$ emission is a combination of pionic and Compton-scattered interstellar light (and subdominant NT bremsstrahlung), *(iii)* the $\lesssim 0.1$ GeV $\gamma$-ray emission is leptonic, and *(iv)* the radio spectrum arises from primary and secondary electron synchrotron emissions at comparable levels. The primary and secondary electron populations are described by a power-law spectrum and a curved spectrum, respectively. Averaged over the set of viable FIR graybody models, the proton spectral index and energy density are $q_p \simeq 2.3$ and $u_p \simeq 385$ eV cm$^{-3}$ (for $n_{\rm H} = 200$ cm$^{-3}$), the (primary) electron and proton maximum energies are $\sim$30 GeV and 7 TeV, respectively, and the magnetic field is $B \simeq 120\,\mu$G. The derived particle and magnetic energy densities are in approximate equipartition.

## 1. Introduction

The central starburst (CSB) regions in intensely star forming galaxies are optimal for the study of relativistic ("cosmic ray"; CR) protons (CRp) and electrons (CRe) that are accelerated by supernova (SN) -driven shocks, and effectively couple to the dense matter, radiation, and magnetic fields. Measurements of the spectral energy distribution (SED) of the nonthermal (NT) radiative yields of CR interactions in CSB provide essential insights into the particle spectra and the physical conditions in star-forming environments.

Gas and CR densities and magnetic and radiation field distributions vary significantly across galaxy disks, generally requiring a spectro-spatial treatment based on a solution to the diffusion-advection equation in cases in which the spatial profiles of relevant key quantities may be specified (e.g., Rephaeli & Sadeh 2019, 2024; Heesen 2021; Strong et al. 2007, 2010; Taillet & Maurin 2003; Wallace 1980; Jones 1978; Syrovatskii 1959). However, in order to gain an insight into spatially averaged NT properties in the relatively small CSB region, we started from the CR yields and deduced the underlying CR steady-state spectra, making it possible (in principle) to deduce the injection spectra.

In previous works (Persic & Rephaeli 2022; Persic et al. 2024) we examined NT emission in nearby galaxies with low-to-moderate star formation rates – the Magellanic Clouds, M31 and M33. Adopting a one-zone model for the extended disk emission, we modeled disk radio emission as a combination of synchrotron and thermal bremsstrahlung and $\gamma$-ray measurements as neutral-pion ($\pi^0$) decay (follow-ing CRp interactions with the ambient gas) and leptonic emission.

In this follow-up study, we consider the local starburst (SB) galaxy M82. The enhanced star formation occurs in the CSB region, which we model as a spheroid of semimajor (semiminor) axis $R_s = 338$ ($h_s = 165$) pc, where the NT emission relevant to this study (see below) is produced. Located at a distance of D = 3.6 Mpc, M82 is seen almost edge-on, with an inclination angle (to the line of sight) of $i \sim 80°$.

A member of the M81 group, M82 shows traces of past tidal encounters. One encounter with M81 less than 400 Myr ago (Hutton et al. 2014), shown by an intergalactic gas bridge, triggered a major SB at a rate of $\sim$10 $M_\odot$ for $\sim$50 Myr. Two subsequent SBs followed, the last of which took place $\sim$ 1 Myr ago (Barker et al. 2008). A prominent galactic wind ("superwind") shoots out of both sides of the galactic disk in nearly perpendicular directions, out to several kiloparsecs, likely driven by the superposed winds of many massive stars and SN explosions. The CSB region is contained in the innermost ($R \lesssim$ 300 pc) disk, surrounded by a molecular ring that has a radius of 400 pc.

M82 emission levels are in accord with the well-known correlations between radio, infrared, and $\gamma$-ray emission for star-forming galaxies (radio–IR: Condon 1992, Bell 2003, Persic & Rephaeli 2007; $\gamma$–IR: Ackermann et al. 2012, Kornecki et al. 2020). [Infrared (IR) emission is variously estimated by the total-IR (TIR) band ($8-1000\,\mu$m), the far-IR (FIR) band ($40-120\,\mu$m), or a specific monochromatic flux (e.g., $60\,\mu$m).] These correlations are traced back to CR acceleration by SN explosions, and the interaction of particles and radiation with the dusty magnetized medium characterizing the sites where massive progenitor stars were formed (Lacki et al. 2010; Bell 2003).

This study aims to determine NT aspects (CR spectra, magnetic field) in the CSB region by spectrally modeling emission in the relevant energy ranges accessible to observations. Radio and $\gamma$-ray spectra from the central region of M82 have long been available (radio: Klein et al. 1988, Seaquist & Odegard 1991, Carlstrom & Kronberg 1991, Williams & Bower 2010, Peel et al. 2011; $\gamma$ ray: VERITAS Collaboration et al. 2009, Abdo et al. 2010). Several studies of the NT emission from the CSB were published, both before and after its $\gamma$-ray detection. Among the former, early estimates (Völk et al. 1989, 1996; Akyüz et al. 1991) and subsequent more detailed models (Persic et al. 2008; De Cea del Pozo et al. 2009) predicted that $\gamma$-ray emission from M82 was detectable by then-current and upcoming detectors. Among the latter, Lacki et al. (2011), Peretti et al. (2019), Krumholz et al. (2020), and Ambrosone et al. (2022, 2024) discussed calorimetric properties and CR physics in the dense CSB environment.

The motivation for the present study stems from an important recent observational development. Diffuse NT X-ray emission was detected in an ellipse of semimajor and semiminor axis $19''.4 \times 9''.4$ ($338\,\mathrm{pc} \times 165\,\mathrm{pc}$) morphologically consistent with the radio and far-infrared (FIR) emission maps of M82's CSB region (Iwasawa et al. 2023). The co-spatial emission in the three bands and the similarity of the NT X-ray and radio-synchrotron spectral indices ($\alpha_{\rm X} \simeq \alpha_{\rm r} \simeq 0.7$) led Iwasawa et al. (2023) to interpret the detected NT X-ray emission as being due to Compton scattering of the FIR photons by radio-emitting CRe. The direct relevance of Compton scattering is obvious given the co-spatial presence in the CSB of CRe and FIR photons. The detected Compton/FIR flux enables a calibration of the CRe spectrum to be made given the obvious relation between Compton emissivity and CRe number density, $j_{\rm Compt} \propto n_e$. Once the CRe spectrum is fully determined, the mean (volume-averaged) magnetic field can be directly deduced. This procedure, which does not rest on equipartition arguments, is possible for the first time in this prototypical SB galaxy.

The paper is organized as follows. In Sect. 2 we review archival data on broadband NT interstellar emission from the CSB of M82, and we summarize our analysis of 16.3 yr of data collected by the Large Area Telescope (LAT) on board the *Fermi* Gamma-ray Space Telescope (fully described in Appendix A). In Sect. 3 we review the radiation fields permeating the CSB region. In Sects. 4 and 5 we describe and discuss our SED modeling. Our conclusions are summarized in Sect. 6.

## 2. Observations of NT emission from the CSB

Diffuse NT emission from M82 has been measured in radio/microwave, X-ray, and $\gamma$-ray (e.g., Gendre et al. 2013; Iwasawa et al. 2023; Ohm 2016), in addition to extensive measurements of diffuse (thermal) emission in the near-ultraviolet, optical, and IR (Hutton et al. 2014; Gurzadyan et al. 2015). In this section we review some of the published observations of NT CSB emission and describe our newly reduced 16.3 years of accumulated *Fermi*-LAT data.

### 2.1. Radio

M82 has been extensively observed over the years (e.g., Kellermann et al. 1969, Klein et al. 1988, Carlstrom & Kronberg 1991, Williams & Bower 2010, Peel et al. 2011). In the present study we focus on emission from the small CSB region from which NT X-ray and $\gamma$-ray emission has been detected. The spectrum of radio emission from this region was presented by Adebahr et al. (2013), based on combined archival Very Large Array (VLA) data (4.996 GHz, 8.327 GHz) and Westerbork Synthesis Radio Telescope (WSRT) data (0.326 GHz and 1.364 GHz). The VLA data (Seacoast & Odegard 1991, Reuter et al. 1992) had previously shown emission mostly confined to the CSB region of M82.

Improved data reduction and instrumental techniques enabled Adebahr et al. (2013) to analyze the radio maps in increased detail. The archival WRST data were reduced by Adebahr et al. (2013) using a new calibration technique able to reach the high dynamical ranges needed to resolve the complex source morphology and map the residual diffuse emission in the central region. The combined datasets yielded total power maps at 0.326, 1.364, 4.996, and 8.327 GHz. At each frequency the intensity was integrated over the angular size of the CSB region (assumed by Adebahr et al. (2013) to have a radius of 450 pc, at a distance of 3.5 Mpc: i.e., $26''.5$). A fifth measurement at 1.67 GHz (by Braun et al. 2007) was included in their power-law (PL) spectral fit by means of free-free absorption by a screen of ambient ionized gas, $S = S_0\,(\nu/\nu_0)^{-\alpha}\,e^{-\tau_{\rm ff}}$, with $\alpha = 0.62 \pm 0.01$ and $\tau_{\rm ff} = 8.2 \times 10^{-2} \nu^{-2.1} {\rm EM}/T_e^{1.35}$, where EM $= (3.16 \pm 0.10) \times 10^5$ cm$^{-6}$pc is the emission measure and $T_e \sim 10^4$K is the warm gas temperature. We adopted the spectral data of Adebahr et al. (2013) after rescaling each flux value by the appropriate factor to match the assumed size of the CSB region from which the NT X-ray emission was measured by Iwasawa et al. (2023). The rescaled data are listed in Table 1.

**Table 1.** CSB radio flux densities.

| Frequency GHz | Flux density Jy | Ref. |
|---|---|---|
| 0.326 | $2.72 \pm 0.37$ | [1] |
| 1.364 | $3.15 \pm 0.20$ | [1] |
| 1.665 | $2.83 \pm 0.19$ | [1] |
| 4.996 | $1.48 \pm 0.09$ | [1] |
| 8.327 | $1.08 \pm 0.06$ | [1] |

[1] Derived from Sect. 4.2 of Adebahr et al. (2013) and rescaled by $(19''.4/26''.5)^2$, where the two angular sizes are the radii of the NT X-ray emission (Iwasawa et al. 2023) and of the "core" radio emission (Adebahr et al. 2013), respectively.

**Table 2.** *Chandra* 5 keV flux density.

| Frequency $\log(\nu/\mathrm{Hz})$ | Flux $10^{-13}\mathrm{erg\,cm^{-2}s^{-1}}$ | Reference |
|---|---|---|
| 18.082 | $1.14 \pm 0.08$ | [1] |

[1] This work, from Iwasawa et al. (2023).

### 2.2. X-ray

X-ray measurements of M82 have shown early on that most of the emission comes from the central region (e.g., Watson et al. 1984), possibly due to either a variable accreting source (Ptak & Griffiths 1999) or a hot ($kT \sim 6-9$ keV) thermal interstellar plasma with a solar metal abundance of $\sim$0.3 (Cappi et al. 1999). The presence of this hot ejecta of many SN explosions, whose superposed shocks drive a superwind into the intergalactic medium, was suggested by Griffiths et al. (2000). Later Strickland & Heckman (2007) found the CSB region to be filled by 6.7 keV He $\alpha$ line emission as well as a marginally significant 6.4 keV Fe K$\alpha$ line and possibly NT continuum.

An improved insight into the origin of X-ray emission came recently from analysis of 570 ks of *Chandra* data by Iwasawa et al. (2023), who found evidence of a NT emission component, at a level of $\sim$70% of the 4−8 keV emission. They deduced that this emission is morphologically consistent with the radio map of the CSB, which they interpreted as arising from Compton scattering of local (i.e., SB-sourced) FIR photons by radio-emitting CRe. This result can be used to estimate the spectral X-ray luminosity at 5 keV: for an energy spectral index of $\alpha = 0.7$ (Iwasawa et al. 2023), the photon spectrum is $N_{\rm ph}(\epsilon) = N_{1\,\rm keV}\,(\epsilon/\mathrm{keV})^{-1.7}$ ph/(cm$^2$ s keV), with energy in kilo-electronvolts. The 4−8 keV spectral energy flux is $\phi_{4-8\,\rm keV} = \int_4^8 \epsilon\, N_{\rm ph}(\epsilon)\,\mathrm{d}\epsilon = 1.17\, N_{1\,\rm keV}$ keV/(cm$^2$ s). Since this should equal $f_{4-8\,\rm keV} = L_{4-8\,\rm keV}/(4\pi D^2)$ erg cm$^{-2}$ s$^{-1}$, where $L_{4-8\,\rm keV} = 0.75 \times 10^{39}$ erg s$^{-1}$ (Iwasawa et al. 2023), it follows that $N_{1\,\rm keV} = 2.21 \times 10^{-4}$ and $N_{5\,\rm keV} = N_{1\,\rm keV}(5/\mathrm{keV})^{-1.7} \simeq 1.43 \times 10^{-5}$. The monochromatic 5 keV energy density flux, $f_{5\,\rm keV} = N_{5\,\rm keV} E_{5\,\rm keV}$, is then (Table 2)

$$f_{5\,\rm keV} \;=\; (1.14 \pm 0.08) \times 10^{-13} \quad \mathrm{erg\,cm^{-2}\,s^{-1}}\,. \qquad (1)$$

**Table 3.** $\gamma$-ray emission: I. *Fermi*-LAT data.

| Frequency $\log(\nu/\mathrm{Hz})$ | Flux $10^{-12}\mathrm{erg\,cm^{-2}s^{-1}}$ | | Ref. |
|---|---|---|---|
| $22.16 \pm 0.08$ | >1.48 | <6.29 | [1] |
| $22.32 \pm 0.08$ | >1.63 | <4.55 | [1] |
| $22.48 \pm 0.08$ | >1.77 | <3.98 | [1] |
| $22.63 \pm 0.08$ | >1.88 | <2.54 | [1] |
| $22.87 \pm 0.16$ | $1.99 \pm 0.19$ | | [1] |
| $23.18 \pm 0.16$ | $1.61 \pm 0.13$ | | [1] |
| $23.50 \pm 0.16$ | $1.78 \pm 0.12$ | | [1] |
| $23.81 \pm 0.16$ | $1.54 \pm 0.13$ | | [1] |
| $24.13 \pm 0.16$ | $1.02 \pm 0.14$ | | [1] |
| $24.44 \pm 0.16$ | $0.94 \pm 0.19$ | | [1] |
| $24.76 \pm 0.16$ | $0.70 \pm 0.23$ | | [1] |
| $25.39 \pm 0.16$ | $0.69 \pm 0.44$ | | [1] |
| $25.70 \pm 0.16$ | $0.53 \pm 0.40$ | | [1] |

[1] This work.

### 2.3. $\gamma$ ray

M82 was measured to be a point source with (non-varying) >200 MeV emission detected at a 6.8 $\sigma$ significance based on 11 months of scanning-mode data since the inception of the *Fermi* mission (Abdo et al. 2010). The measured emission was interpreted as originating from the interaction of relativistic and ambient protons, suggesting a link between massive star formation and $\gamma$-ray emission. This interpretation was found to be valid for star-forming galaxies in general (Ackermann et al. 2012; also Mannheim et al. 2012 and Chen et al. 2024). Here we analyze 195.6 months (16.3 yr) of scanning-mode LAT data (Table 3), detailed in Appendix A.

Based on 137 hr of high-quality Very Energetic Radiation Imaging Telescope Array System (VERITAS) data, collected in 2008-2009, a steady >0.7 TeV flux of $(3.7 \pm 0.8_{\rm stat} \pm 0.7_{\rm syst}) \times 10^{-13}$ cm$^{-2}$ s$^{-1}$ was detected (at a 4.8 $\sigma$ significance level) from M82 (VERITAS Collaboration et al. 2009) − the first detection of a SB galaxy at tera-electronvolt energies. The differential photon $\gamma$-ray spectral index was $\Gamma = 2.5 \pm 0.6_{stat} \pm 0.2_{syst}$. Owing to a follow-up observational campaign in 2011−2022, 254 hours of good-quality data are now available. The expanded dataset and improved analysis techniques yielded a significant (6.5$\sigma$) point-like detection (VERITAS Collaboration et al. 2025). The resulting flux is F(>450 GeV) = $(3.2 \pm 0.6_{\rm stat} \pm 0.6_{\rm syst}) \times 10^{-13}$ cm$^{-2}$ s$^{-1}$ ($\sim$0.4% of the Crab Nebula flux above the same threshold), with a spectral slope of $\Gamma = 2.3 \pm 0.3_{stat} \pm 0.2_{syst}$, and no obvious spectral cutoff at energies $\lesssim$5 TeV. The 2008−2022 VERITAS data are summarized in Table 4 and Appendix B. In summary, the broadband $\gamma$-ray SED of M82 used in the present study extends from 50 MeV to 16.25 TeV, with data taken during the periods 2008−2022 (VERITAS) and 2008−2024 (*Fermi*-LAT).

**Table 4.** $\gamma$-ray emission: II. VERITAS data.

| Frequency $\log(\nu/\mathrm{Hz})$ | Flux, +err, −err $10^{-12}\mathrm{erg\,cm^{-2}s^{-1}}$ | | | Ref. |
|---|---|---|---|---|
| $26.01 \pm 0.08$ | $-12.43,$ | $+0.24,$ | $-0.58$ | [1] |
| $26.18 \pm 0.08$ | $-12.64,$ | $+0.19,$ | $-0.36$ | [1] |
| $26.34 \pm 0.08$ | $-12.60,$ | $+0.14,$ | $-0.20$ | [1] |
| $26.51 \pm 0.08$ | $-12.59,$ | $+0.13,$ | $-0.18$ | [1] |
| $26.68 \pm 0.08$ | $-12.57,$ | $+0.12,$ | $-0.17$ | [1] |
| $26.84 \pm 0.08$ | $-12.91,$ | $+0.23,$ | $-0.52$ | [1] |
| $27.01 \pm 0.08$ | $-12.82,$ | $+0.21,$ | $-0.41$ | [1] |
| $27.18 \pm 0.08$ | $<-12.56$ | | | [1] |
| $27.34 \pm 0.08$ | $<-12.65$ | | | [1] |
| $27.51 \pm 0.08$ | $<-12.73$ | | | [1] |

[1] VERITAS Collaboration et al. (2025).

## 3. Radiation fields

A detailed description of the ambient radiation fields is needed to predict $\gamma$-ray emission from Compton scattering of target photons by the radio-emitting CRe. The ambient radiation fields are extragalactic and local. The former include the cosmic microwave background (CMB), a Planckian with a temperature of $T_{\rm CMB} = 2.735\,\mathrm{K}$ and energy density of $u_{\rm CMB} = 0.25$ eV cm$^{-3}$, and the extragalactic background light (EBL). The EBL is the superposed emission from direct (optical) and dust-reprocessed (into IR) starlight integrated over the star formation history of the Universe. Its spectrum has two humps, optical and IR, peaking at $\sim$1 $\mu$m and $\sim$100 $\mu$m. These humps are described as diluted Planckians, characterized by a temperature, $T$, and a dilution factor, $C_{\rm dil}$. [1] We adopted the commonly used EBL model by Franceschini & Rodighiero (2017), numerically approximated as a combination of diluted Planckians (Eq. 1 of Persic et al. 2024).

The dominant radiation field in the CSB region arises from the local stellar population. Similar to the EBL in shape and origin, it is dominated by the IR and optical humps. Full-band luminosities of these thermal components are needed to determine $n(\epsilon)$, the spectral distributions of target IR/optical photons (of energy $\epsilon$) that are Compton-upscattered to X-ray/$\gamma$-ray energies. We proceeded as follows:

*(i)* The total optical-band luminosity was computed from the narrow B-band luminosity ($B$=8.76, $B-V$=1.15; Blackman et al. 1979) by applying a suitable bolometric correction, $B_{\rm bol} = B + BC_B$ with $BC_B = -0.85 - (B - V)$ (Buzzoni et al. 2006; Persic & Rephaeli 2019). The corresponding number density spectrum of optical photons is a (diluted, unmodified) Planckian. The optical energy density inside the source (assumed to be homogeneous), calculated from $u_{\rm opt} \simeq (9/4) L_{\rm IR;\,opt}/[4\pi R_s^2 c]$ (Ghisellini 2013) [2], is $u_{\rm opt} \sim 6.8 \times 10^{-10}$ eV cm$^{-3}$. The optical-hump temperature, estimated applying Wien's displacement law to the optical-peak frequency, 1.5 $\mu$m (Vaccari & Franceschini 2008), is $T_{\rm opt} = 2000\,\mathrm{K}^{[c]}$. The corresponding dilution factor is $\log(C_{\rm dil}^{\rm opt}) = -8.248$.

*(ii)* The total-infrared (TIR) energy density was derived from $u_{\rm FIR} \sim 8.2 \times 10^{-10}$ eV cm$^{-3}$ (Iwasawa et al. 2023), multiplied by 1.65 to convert FIR to TIR fluxes (Persic & Rephaeli 2007), and further corrected for the mean photon-escape time in a (quasi) spherical homogeneous source (see above). This yielded $u_{\rm TIR} \sim 3\times10^{-9}$ erg cm$^{-3}$. The calculation of the IR dilution factor, $C_{\rm dil}^{\rm IR}$, was less straightforward than in the optical case because the ambient IR emission is clearly not blackbody but "graybody." The calculation is outlined below.

◊ *IR graybody emission*

The spectral flux from a dusty cloud is described as a modified blackbody (graybody), i.e., the product of a Planckian function, $B(\nu, T_{\rm d})$, by the dust emissivity function, $Q(\nu) = 1 - e^{-\tau_{\rm gb}}$, where $\tau_{\rm gb} = (\nu/\nu_0)^\beta$ is the dust-related optical depth (Hildebrand 1983) and $0 < \beta \lesssim 2$ is the dust emissivity index (e.g., Yun & Carilli 2002) [3]. The reference frequency, $\nu_0$, corresponds to $\tau_{\rm gb} = 1$. At frequencies of $\nu \gg \nu_0$, the dust spectrum is optically thick and reduces to a blackbody, whereas at $\nu \ll \nu_0$ it describes an optically thin graybody, $j_\nu(\beta, T_{\rm d}) = B(\nu, T_{\rm d})\,(\nu/\nu_0)^\beta \propto \nu^{2+\beta} T_{\rm d}$ (which reduces to the Rayleigh-Jeans law for $\beta = 0$).

The IR emission of M82 is represented by an optically thin graybody, with $\nu_0$ corresponding to the peak emission frequency, $\nu_0 = \nu_p = 2.5 \times 10^{12}$ Hz (Yun & Carilli 2002; Peel et al. 2011). In evaluating Compton scattering of IR photons by CRe in the CSB we assume that the spectral energy density of target photons is $n_{\rm IR}(\epsilon) = 8\pi/(h^3c^3)\,\epsilon^2/(e^{\epsilon/k_B T_{\rm d}} - 1)(\epsilon/\epsilon_0)^\beta$ (which replaces Eq. A.7 of Persic et al. 2024). As for the $(\beta, T_{\rm d})$ pair, whose combination in the $\nu^\beta B(\nu, T_{\rm d})$ (graybody) fit describes the dust spectral energy profile (e.g., Fig. 1 of Yun & Carilli 2002), we used several published values relative to various FIR datasets [4] that quantify the overall FIR modeling uncertainty. For these $(\beta, T_{\rm d})$ pairs, Eq. (12) of Elia & Pezzuto (2016) specifies values for the peak emission frequency, $\nu_p \sim 2.5\times10^{12}$ Hz, consistent with the observed value (e.g., Peel et al. 2011). In principle these $\nu^\beta\,B(\nu, T)$ fits are mutually consistent; nevertheless, we present SED models for all of them to check the specificity of the NT SED versus the IR SED.

To determine the appropriate dilution factor for an optically thin graybody, we proceeded as follows. The power radiated by an optically thin graybody per unit surface area, $W_{\rm GB} = \int_0^\infty (\frac{\nu}{\nu_0})^\beta\, B_\nu(T)\,{\rm d}\nu$, is

$$W_{\rm GB} \;=\; \frac{2\pi k_B^{4+\beta}}{h^{3+\beta}c^2}\,\frac{1}{\nu_0^\beta}\,\Gamma(4+\beta)\,\zeta(4+\beta)\;T^{4+\beta} \tag{2}$$

[1] The dilution factor is the ratio of the actual energy density, $u$, to the energy density of an undiluted blackbody at the same temperature, $T$, i.e., $C_{\rm dil} = u/(\frac{4}{c}\sigma_{\rm sb}T^4)$, where $\sigma_{\rm sb}$ is the Stefan-Boltzmann constant.

[2] The factor 9/4 accounts for the mean photon escape time in the CSB (quasi) spherical geometry.

[3] Analytical relations between graybody parameters are provided by Elia & Pezzuto (2016).

[4] The adopted $(\beta, T_{\rm d})$ pairs are: 1.0, 48 K (Colbert et al. 1999: (43−197) $\mu$m, *Infrared Space Observatory* Long Wavelength Spectrometer data); 1.3, 48 K (Hughes et al. 1994: 40$\mu$m−3mm, James Clerk Maxwell Telescope (JCMT) and National Radio Astronomy Observatory on Kitt Peak (NRAO) data); 1.5, 47 K (Hughes et al. 1990: 10$\mu$m−3mm, JCMT and archival data; and Klein et al. 1988: 10.7−32 GHz Effelsberg 100-m telescope and archival data); 1.65, 32 K and 2.1, 25 K (Peel et al. 2011: (28.5−857) GHz, *Planck* and *Wilkinson Microwave Anisotropy Probe* data); and 2.0, 30 K (Thronson et al. 1987: 40$\mu$m−1.3mm, NRAO and archival data).

**Table 5.** Interstellar medium parameters.

| $n_{\rm HI}$ cm$^{-3}$ | $n_{\rm H_2}$ cm$^{-3}$ | $n_i$ cm$^{-3}$ | $Z$ | EM cm$^{-6}$pc | $T_e$ K | $F({\rm H}\alpha)$ erg/(cm$^2$s) |
|---|---|---|---|---|---|---|
| 10[a] | 65[b] | 60[c] | 1[d] | 3 E+5 [e] | $10^4$ | 7.9 E−11 [f] |

[a]Lo et al. (1987); [b]Weiss et al. (2001), De Cea del Pozo et al. (2009); [c]Houck et al. (1984); the hot ionized gas contributes $n_e = 0.1$ cm$^{-3}$ (Iwasawa et al. 2023); [d]Iwasawa et al. (2023); [e]model of a warm-ionized-gas screen absorbing a synchrotron source (Adebahr et al. 2013); [f]Kennicutt et al. (2008).

where $\Gamma(z)$ and $\zeta(z)$ are Euler's gamma function and Riemann's zeta function, respectively (see Eqs. [30],[31] of Elia & Pezzuto 2016).

The graybody equivalent of the blackbody displacement law is $\nu_p \simeq \frac{k_B T}{h}(3+\beta)$ (Elia & Pezzuto 2016, Eq. 12). If $\nu_0 = \nu_p$, then

$$W_{\rm GB} \;=\; \frac{2\pi k_B^4}{h^3 c^2}\,\frac{1}{(3+\beta)^\beta}\,\Gamma(4+\beta)\,\zeta(4+\beta)\;\; T^4 \;= \;=\; \sigma^{\rm GB}(\beta)\;\; T^4\,, \tag{3}$$

where we introduced the quantity

$$\sigma^{\rm GB}(\beta) \;\equiv\; \frac{2\pi k_B^4}{h^3 c^2}\,\frac{1}{(3+\beta)^\beta}\,\Gamma(4+\beta)\,\zeta(4+\beta)\,. \tag{4}$$

A graybody reduces to a blackbody for $\beta = 0$, i.e.,

$$\sigma^{\rm GB}(0) \;=\; \sigma_{\rm sb} \;=\; \frac{2\pi k_B^4}{h^3 c^2}\,\Gamma(4)\,\zeta(4) \tag{5}$$

where $\sigma_{\rm sb}$ is the Stefan-Boltzmann constant. Thus Eq. (4) can be written as

$$\sigma^{\rm GB}(\beta) \;=\; \frac{1}{(3+\beta)^\beta}\,\frac{\Gamma(4+\beta)\,\zeta(4+\beta)}{\Gamma(4)\,\zeta(4)}\;\sigma_{\rm sb}\,. \tag{6}$$

Equation (6) yields the substitute to the Stefan-Boltzmann constant for an optically thin GB specified by the spectral index, $\beta$, with $\nu_0 = \nu_p$. The latter is generally valid in SB galaxies (Yun & Garilli (2002)). Equation (6) is a monotonically decreasing function of $\beta$; clearly, $\lim_{\beta\to 0}\sigma^{\rm GB}(\beta) = \sigma_{\rm sb}$ and $\lim_{\beta\to\infty}\sigma^{\rm GB}(\beta) = 0$. Indeed, $\sigma^{\rm GB} \approx 0$ already at $\beta = 10$. The dilution factor clearly depends on the parameter pair $(\beta, T)$ describing the graybody, $C_{\rm dil}^{\rm IR} = u_{\rm IR}/[\frac{4}{c}\sigma^{\rm GB}(\beta)\,T^4]$ (specified below) [5].

## 4. SED modeling

We aim to determine NT quantities (CR, magnetic field) in the CSB region by comparing the SED dataset to the emission predicted by relevant radiative processes (well-known formulae are collected in Appendix A of Persic et al. 2024). Given the relatively small size of the CSB and lacking any spatial information on NT emission from this region, we assumed the CR spectral distributions to be time-independent, locally isotropic, and uniform.

Below are the CR spectral distributions:

- The CRp spectrum is a PL in energy, $N_p(E_p) = N_{p,0} E_p^{-q_p}$ cm$^{-3}$ GeV$^{-1}$ for $m_p c^2 < E_p < E_p^{\rm max}$ (energies in gigaelectronvolts). The normalization and slope, $N_{p,0}$, $q_p$, are free parameters.
- Secondary CRe are produced from $\pi^\pm$ decays following p−p interactions of CRp with the ambient gas. The source spectrum of freshly created secondary CRe, essentially a PL in energy, is modified by energy losses: radiative losses (Coulomb, bremsstrahlung, synchrotron/Compton) and, in the present case, also superwind advection losses (Longair 1981),

$$b_{\rm adv} \;=\; 2.16\times 10^{-17}\,\gamma\,\left(\frac{V_{\rm wind}}{\rm km/s}\right)\left(\frac{r_h}{\rm kpc}\right)^{-1}\;\;{\rm s}^{-1} \tag{7}$$

With the CRp spectrum and the environmental parameters (density, magnetic field, superwind speed) in the CSB region specified (see Tables 5 and 6), the secondary electron spectrum has no free parameters. For computational ease, the secondary spectrum can be analytically approximated as in Eq. (A.29) of Persic et al. (2024).
- The primary CRe spectrum is a PL in Lorentz factor $\gamma$, $N_e(\gamma) = N_{e,0}\gamma^{-q_e}$ cm$^{-3}$ (per unit $\gamma$) for $\gamma_{\rm min} < \gamma < \gamma_{\rm max}$ (with $\gamma_{\rm min} = 100$). The normalization and slope, $N_{e,0}$ and $q_e$, are free parameters.

Our modeling procedure began with fitting a pionic emission profile to the high-frequency ($\nu \gtrsim 10^{24.5}$ Hz) $\gamma$-ray data with free normalisation, a slope, and a high-end cutoff; photon−photon opacity is not important, as is discussed in Sect. 6 and shown in Fig. 3. The emission spectrum from $\pi^0$-decay has constraining power owing to its characteristic "shoulder" at $\sim$100 MeV; if apparent in the data at higher energies, a spectral cutoff corresponds to $E_p^{\rm max}$. For the set of models developed in this study, corresponding to different ($\beta$, $T_{\rm d}$) pairs (see Sect. 5), we deduced CRp spectral slopes of $q_p \simeq 2.3$, energy cutoffs of $E_p^{\rm max} \approx 7$ TeV, and energy densities of $u_p \simeq 385$ eV cm$^{-3}$ (Table 6).

Next we used the secondary- and primary-CRe spectra to calculate the monochromatic 5 keV Compton/starlight emission. With the secondary CRe spectrum fully determined, unlike the primary CRe spectrum, which was assumed, fitting the 5 keV point allowed us to determine $N_{e,0}$. We assumed a primary CRe injection index of $q_e \leq 2.2$. This assumption was based on observational evidence that the distribution of radio synchrotron indices for Galactic SN remnants (SNRs), i.e., the putative sites of Galactic CR acceleration, peaks at $0.50 \lesssim \alpha_r \lesssim 0.55$ (Klein et al. 2018), i.e. $2 \lesssim q_e \lesssim 2.1$. While there are many SNRs in the CSB region, their filling factor in the CSB is low, so the CRe index averaged over the region must be steeper than that measured locally in SNRs.

Radio emission was modeled as CRe synchrotron in a disordered (spatially averaged) magnetic field, $B$. Deducing $B$ directly from spectral modeling, rather than assuming particle-field energy equipartition, is a much-needed development in NT spectral analyses of M82 and other SB galaxies. The lack of an obvious high-$\nu$ cutoff in the radio spectrum makes it impossible to evaluate $\gamma_{\rm max}$ from radio data alone; however, the Compton/starlight $\gamma$-ray emission suggests values of $\gamma_{\rm max} \sim 5.5\times 10^4$. At high radio frequencies, the $\nu^{-0.1}$ component represents diffuse thermal free-free emission from a warm ($\sim 10^4$ K) ionized plasma (e.g., Spitzer 1978), possibly concentrated in the CSB (Shopbell & Bland-Hawthorn 1998). This emission may be gauged by the H$\alpha$ flux, $F(\mathrm{H}\alpha)$, if both come from the same HII regions, since in this case the relevant warm-plasma parameters (temperature, density, filling factor) are the same. In this case the measured $F(\mathrm{H}\alpha)$ can be used to predict the level of free-free emission. To do this we used Eq. (17) of Klein et al. 2018 with a warm-plasma temperature of $T_e = 10^4$ K and (aperture-integrated) flux of $F(\mathrm{H}\alpha) = (0.79 \pm 0.05) \times 10^{-10}$ erg cm$^{-2}$ s$^{-1}$ (Kennicutt et al. 2008). The radio spectral data and model are shown in Fig. 2.

[5] $\log C_{\rm dil}^{\rm IR} = -1.081, -1.056, -1, -0.315, -0.159$, and $0.172$, for the sequence of $(\beta, T_{\rm d})$ pairs reported in Table 6.

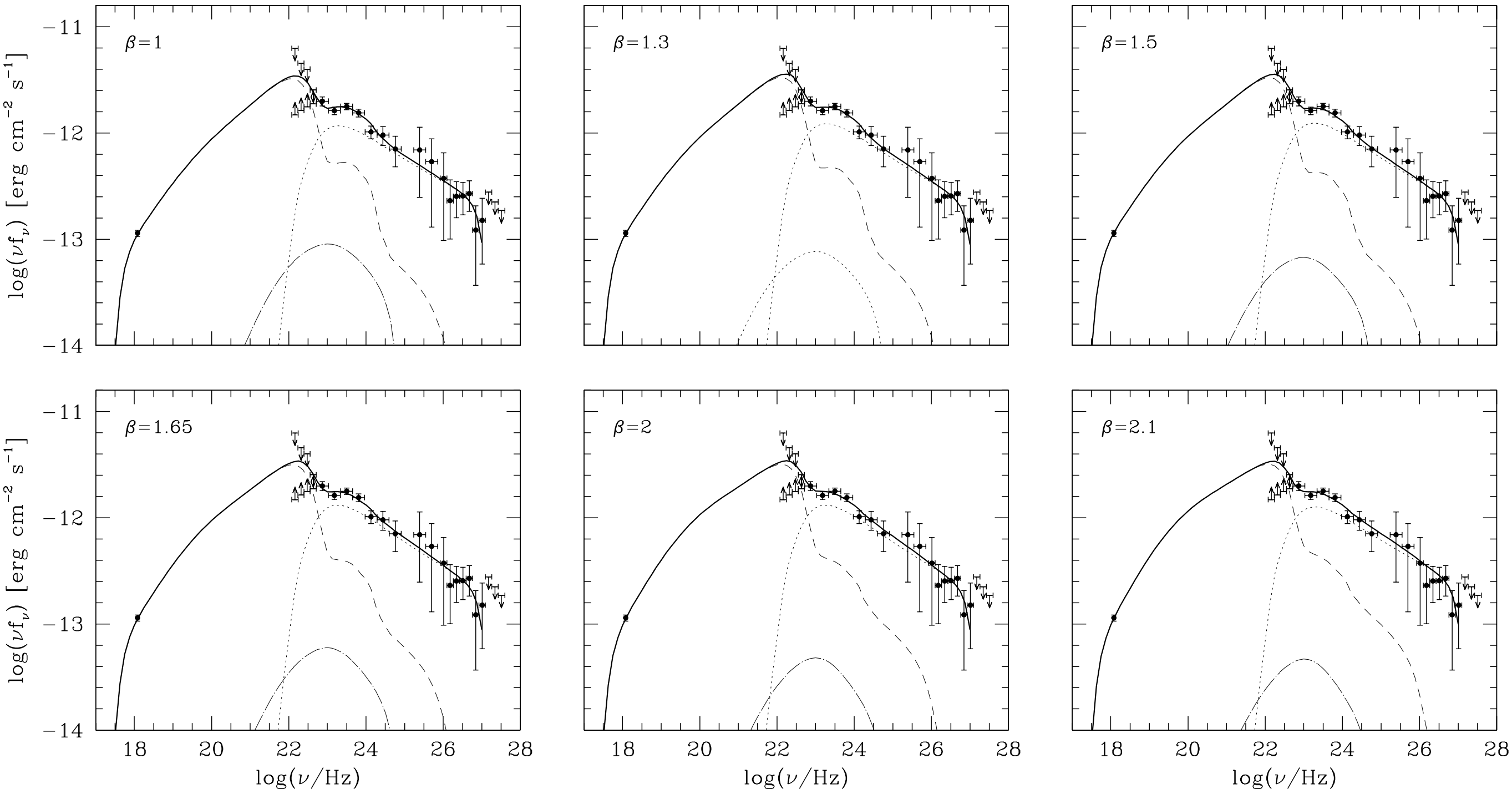

**Fig. 1.** X-ray/$\gamma$-ray range of the CSB SED. The emission model (thick solid curve), including total (i.e., primary plus secondary) Compton/M82-EBL-CMB radiation (dashed curve) and NT bremsstrahlung (dash-dotted curve), and a pionic component (dotted curve), is overlaid with data from Tables 2–4 (black dots). The leptonic component is dominated by the Comptonization of the local FIR radiation. The 5 keV flux, resulting from the NT 4–8 keV flux measured by Iwasawa et al. (2023), is fit by (basically) the local Compton/FIR yield, in which the secondary electron spectrum is uniquely set by the CRp spectrum fit to the $\gamma$-ray data and the primary-electron spectrum is normalized to match the 5 keV flux.

As in our (NT) analyses of M31 and M33 (Persic et al. 2024), we consider thermal and NT electrons to be homogeneously mixed throughout the CSB region (Förster Schreiber et al. 2001; Seaquist et al. 1985). The emerging intensity is $I_\nu \propto j_\nu(1 - e^{-\tau_{ff}(\nu)})/\tau_{ff}(\nu)$, where $j_\nu$ is the (synchrotron plus thermal free-free) spectral emissivity and $\tau_{ff}(\nu)$ is the optical depth for free-free absorption by thermal plasma (Persic et al. 2024, Eqs. 2 and 3). We consider this "mixed absorption-emission" geometry to be more realistic then the "screened source" geometry adopted by Adebahr et al. (2013), who considered a synchrotron source absorbed by a foreground screen of $10^4$ K plasma. Our model matches the observed radio spectrum equally well by assuming EM $\approx 0.75 \times 10^6$ cm$^{-6}$pc, which implies that $< n_e^2 >^{1/2} \approx 40$ cm$^{-3}$, consistent with the derived range of free electron densities. [6]

Having determined the CRe spectra, the full NT bremsstrahlung and Compton-scattered starlight $\gamma$-ray yields were calculated using standard emissivity formulae. The total radiative yields from CRe interactions were added to the pionic yield to fit the full *Fermi*-LAT range of $\gamma$-ray data (Fig. 1). An acceptable fit was obtained iteratively by varying the parametrized proton spectrum. We find that $\gamma$-ray emission is dominated by the pionic component over the high-frequency LAT and VERITAS frequency range, but Compton-scattered starlight dominates emission at the lowest LAT energies (50−100 MeV).

## 5. Results and discussion

Different FIR datasets analyzed by various authors have resulted in a range of $(\beta, T_\mathrm{d})$ values. This observational uncertainty in the parameters of the FIR graybody model for the dusty CSB region results in a range of viable SED models presented in the previous section. We note that the models listed in Table 6 cover the observational range of values of the parameters $\beta$ and $T_\mathrm{d}$; clearly, their respective statistical

[6] Densities of thermal electrons were derived from doubly ionized sulphur, neon, and argon forbidden-line ratios (Förster Schreiber et al. 2001). Observations of these lines sample optically unseen ionized gas and provide a way to directly estimate the average electron density, $n_e$. The [S III] ratio is the most sensitive at low $n_e$ because the upper level of the transitions have the lowest critical densities. For example, the [S III, 18.7$\mu$m/33.5$\mu$m] line ratio, taken over a central region of M82 encompassing the central SB, implies $n_e \sim 120^{+500}_{-120}$ cm$^{-3}$ considering a range of plasma temperatures, $T_e \sim 5000$–$10^4$ K (Houck et al. 1984): specializing this analysis to the "standard" warm ionized gas temperature of $T_e = 10^4$ K, the resulting density is $n_e \sim 60$ cm$^{-3}$. For definiteness, in this paper we assume that $n_e = 60$ cm$^{-3}$ in the central SB.

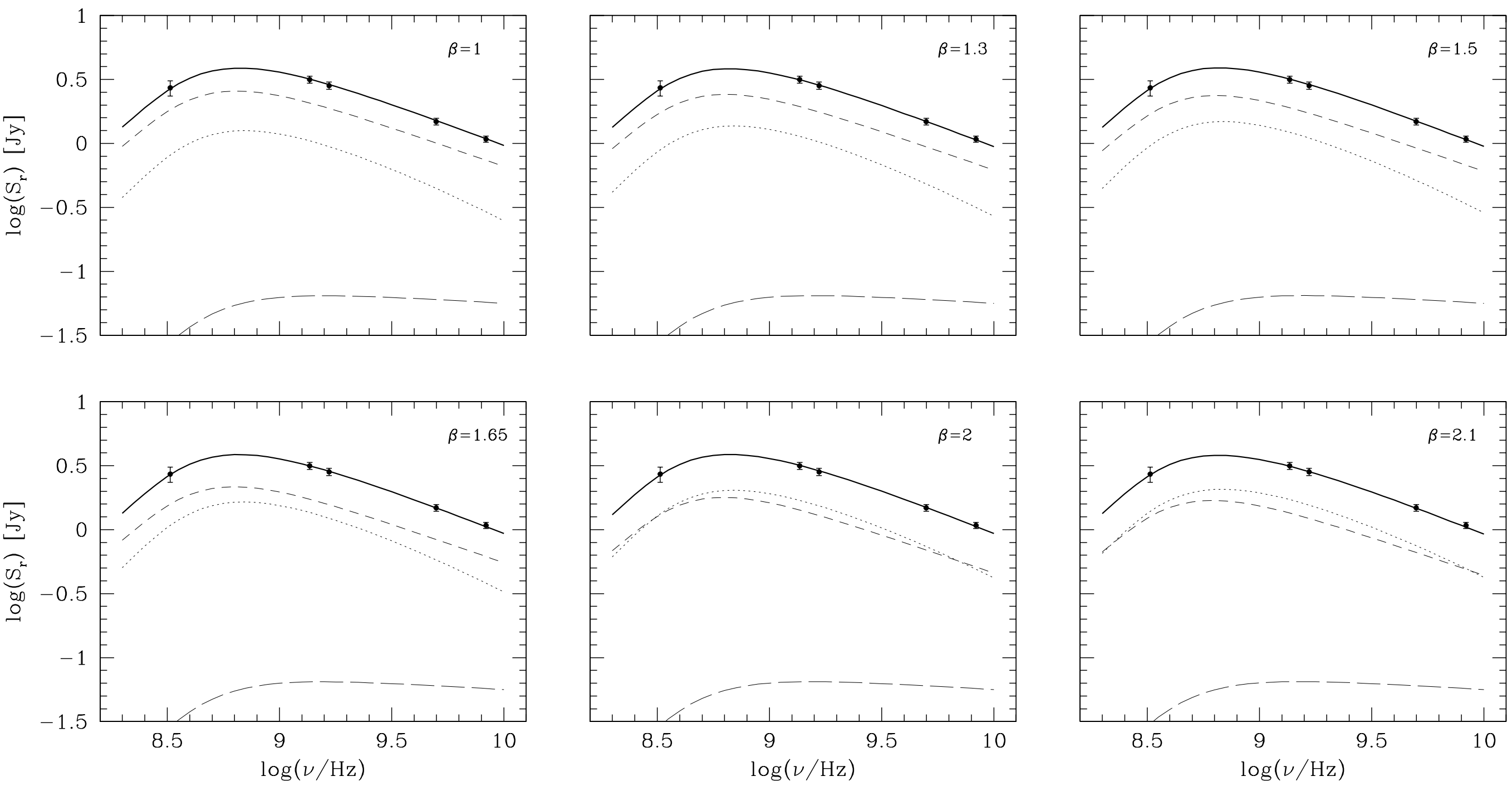

**Fig. 2.** Model radio spectra for the values of $\beta$ considered in this study (thick solid curves), each including primary and secondary electron synchrotron emission (short-dashed and dotted curves, respectively) and thermal free-free radiation (long-dashed curves), are overlaid with data from Table 1 (black dots).

**Table 6.** SED model parameters.

| dust model<br>$\beta$; $T_{\rm d}$ | $q_e$ | $\gamma_{\rm max}$<br>$10^4$ | $u_e^{[a]}$ | $q_p$ | $E_p^{\rm max}$<br>TeV | $u_p^{[a]}$ | $q_1$ | $q_2$ | $\gamma_{b1}$<br>$10^3$ | $\gamma_{b2}$<br>$10^5$ | $\eta$ | $u_{se}^{[a]}$ | $B$<br>$\mu$G | $B$<br>$B_{\rm eq}$ | EM[b] | $\chi^{2\,[c]}_{\nu}$ |
|---|---|---|---|---|---|---|---|---|---|---|---|---|---|---|---|---|
| 1.00; 48K | 2.2 | 5.7 | 5.87 | 2.27 | 7 | 360 | 0.14 | 3.39 | 1.6 | 1 | 0.2 | 1.51 | 97 | 0.80 | 0.78 | 0.95 |
| 1.30; 48K | 2.2 | 5.7 | 4.77 | 2.28 | 7 | 375 | 0.10 | 3.39 | 1.5 | 1 | 0.2 | 1.42 | 106 | 0.86 | 0.77 | 0.90 |
| 1.50; 47K | 2.2 | 5.6 | 4.03 | 2.28 | 7 | 380 | 0.09 | 3.39 | 1.4 | 1 | 0.2 | 1.34 | 116 | 0.93 | 0.76 | 0.91 |
| 1.65; 32K | 2.2 | 5.5 | 3.29 | 2.29 | 7 | 400 | 0.09 | 3.39 | 1.4 | 1 | 0.2 | 1.34 | 124 | 0.97 | 0.75 | 1.01 |
| 2.00; 30K | 2.19 | 5.4 | 2.16 | 2.29 | 7 | 400 | 0.09 | 3.39 | 1.4 | 1 | 0.2 | 1.34 | 142 | 1.11 | 0.72 | 0.99 |
| 2.10; 25K | 2.17 | 5.2 | 1.65 | 2.27 | 7 | 390 | 0.09 | 3.39 | 1.4 | 1 | 0.2 | 1.30 | 143 | 1.14 | 0.70 | 1.05 |

[a] eV cm$^{-3}$; [b] $10^6$ cm$^{-6}$pc. [c] $\nu = 9$ degrees of freedom.

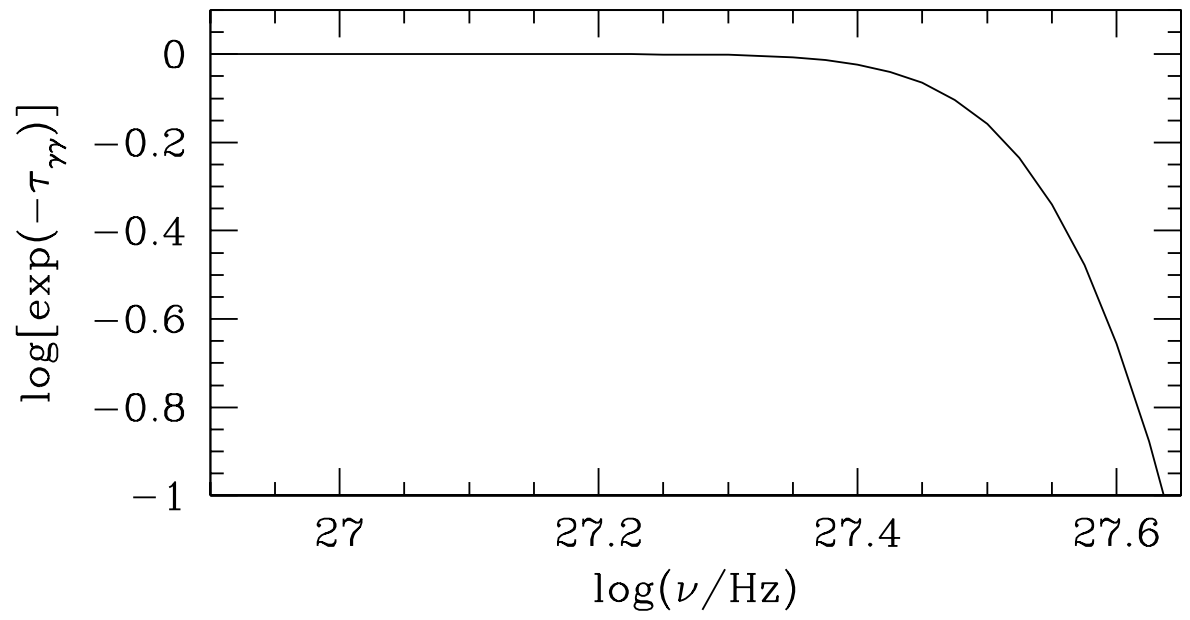

**Fig. 3.** Photon–photon absorption, described as $\exp(-\tau_{\gamma\gamma})$.

likelihoods (quantified by the value of $\chi^2_\nu$) do not imply a relative (model) preference.

Our SED models share basic similarities but also differ significantly, as is briefly discussed below. The models are similar in that the $\gamma$-ray spectrum is viably modeled as pionic emission for most of the measured range except at the lowest LAT energies, where the peak of the Compton-scattered intense local FIR radiation field dominates. The monochromatic 5 keV flux lies on the rising branch of such Compton/FIR starlight. Thus the CRe and CRp spectra are anchored to the X-ray and high-energy $\gamma$-ray emissions, respectively. The radio spectrum is synchrotron emission by primary and secondary CRe in comparable contributions (plus a very minor level of thermal bremsstrahlung), so the volume-averaged magnetic field is directly linked to the radio spectrum.

Model differences stem from the fact that, for a given $T_{\rm d}$, different values of $\beta$ imply different densities of the FIR target photons that are Compton-scattered by the radio electrons into the X-ray/$\gamma$-ray domain. This is exemplified by the models having $\beta = 1, 1.3, 1.5$: a higher $\beta$ implies a higher Compton/(CMB+starlight) emission, so for consistency with the 5 keV flux constraint, the associated primary CRe spectrum must be renormalized lower. Consequently, the deduced $B$ has to be higher in order for the computed synchrotron emission to match the radio data (Table 6).

Our treatment is based on the attainment of a steady state in the CSB region; namely, that particle acceleration and energy loss rates are roughly comparable during the SB phase. Supporting evidence for the validity of this assumption can be seen from the (deduced) near-equipartition between CR and magnetic energy densities, by the appreciable role of secondary CRe in providing the requisite complementary radio synchrotron and X-ray/$\gamma$-ray Compton/FIR emission, and – more generally – by the effectiveness of an isotropic, time-independent modeling of the SED data.

Clearly, this study focuses on the radiative manifestations of NT degrees of freedom in the dense magnetized, likely turbulent CSB gas. As such, it is obviously quite distinct from the more micro-physically detailed (but necessarily parametrized) MHD studies of the strongly coupled media in the CSB. Insights gained from the latter could possibly have some relevance for CR propagation modes and the small-scale morphology of magnetic fields. Given that our analysis is merely spectral (not spectro-spatial), it only relates to (what are essentially) the coarse-grained quantities that characterize the gas, NT particles, and magnetic field. Insights from the results of our spectral analysis could be useful in limiting the parameter space of the more comprehensive MHD modeling of turbulent gas in gas-rich environments, such as in CSB regions, and throughout galactic disks (e.g. Poggianti et al. 1999; Bland-Hawthorn et al. 2024, 2025). In the following we discuss key aspects of our analysis and its results.

• *Lepto-hadronic interplay.* Although the $\gamma$-ray spectrum is naturally interpreted as a pionic emission at $\log(\nu) \gtrsim 25$, a leptonic component (i.e., Compton-scattered starlight [7] ) is necessary to model LAT data at lower frequencies. The interplay between leptonic and hadronic emissions in the model is best revealed at $\log(\nu) \sim 22.7$, where the two components crossover: Compton-scattered starlight declines steeply from its peak at $\log(\nu) \sim 22$ and pionic emission rises steeply to its own peak at $\log(\nu) \sim 23.25$ (Fig. 1). At about the same frequency, $\log(\nu) \sim 23.5$, the Compton/optical peak enhances the hadronic peak. As a consequence the model SED exhibits a local minimum between the leptonic and the hadronic peaks.

The LAT good-quality (spectrally resolved) data in the range $\log(\nu) = 22.7-24$ of the pionic hump enable the two different spectral components to be separated. However, at lower frequencies (i.e. $\lesssim 200$ MeV) just below this range the data points are upper and lower limits, and thus can only bracket the viable range of the model SED. With the predicted (fully determined) Compton/FIR shape of the SED at energies just below $\sim 207$ MeV ($\log(\nu) = 22.7$), it seems likely that the last two upper limits must closely trace the SED in order for the latter to hit the first statistically sound flux point (where the upper and lower limits converge), at $\log(\nu) = 22.87$. Should the curve be even slightly lower, it would violate the quickly rising lower limits. Thus this highly structured SED curve and the arrowhead pattern of $\lesssim 200$ MeV limits converging to the first statistically detected flux point enable the (model-based) conclusion that these limits meaningfully constrain the SED model, so much so that the LAT-data SED can be safely traced down to $\log(\nu) = 22.4$ (i.e. $\sim$100 MeV). Higher-sensitivity measurements are needed to better constrain the emission in the crucial 10−200 MeV band; this could be achieved by a dedicated mega-electronvolt–giga-electronvolt orbiting telescope such as e-Astrogam (De Angelis et al. 2017; Rando et al. 2019).

• *Uncertainty in the level of optical luminosity.* The largest uncertainty in the intrinsic luminosity of the CSB is on the optical emission, i.e., the level of de-reddening necessary to recover the intrinsic optical flux from the measured one. The lower the de-reddening – i.e., the dimmer the deduced intrinsic flux of the optical hump – the less prominent the Compton/optical hump (at $\log(\nu/\mathrm{Hz}) \simeq 23.5$): as a consequence, at $\log(\nu/\mathrm{Hz}) \approx 23.3$ our SED models would no longer exhibit a local minimum (which is indeed shown by the LAT data) but rather an inflection point. Thus, adopting a lower optical luminosity would worsen the match between model and data in this frequency range. In summary, the secondary peak shown by the LAT data at $\log(\nu/\mathrm{Hz}) \approx 23.5$ suggests a relatively high intrinsic optical luminosity, $L_B \gtrsim 7.7 \times 10^{43}$ erg s$^{-1}$, similar to that derived from Blackman et al. (1979) who argued for a high color index – especially in the dusty CSB. As the optical emission of the optical component is described as a blackbody (the graybody model only applies to the FIR emission), its dilution factor is the standard blackbody one (see fn. 6).

• *CRp.* Our fitting CRp spectrum has an index, $q_p \approx 2.3$, consistent with the one derived by the VERITAS Collaboration et al. (2025). The limiting CRp energy resulting from our fits, $E_p^{\max} \approx 7$ TeV, is comparable to the maximum energies attained by CR accelerated by shocks in a uniform medium (with a mild dependence on density for $n_\mathrm{H} = 1 \div 10^2$ cm$^{-3}$) during the Sedov-Taylor phase of SN evolution. Such energies are an outcome of semi-analytical models of particle acceleration, which are based on kinetic simulations for a wide range of astrophysical shocks and account for the interplay between particle acceleration, magnetic-field amplification, and shock evolution (Diesing 2023; Bell et al. 2013).

The CRp energy density, $u_p$, in our models is in the range of 360−400 eV cm$^{-3}$ (for $n_\mathrm{H} = 200$ cm$^{-3}$). An independent estimate of $u_p$ may provide a consistency check. Combining the SN frequency, $\mathcal{R}_\mathrm{SN}$, with the CRp residency timescale, $\tau_\mathrm{res}$, in the source volume, $\mathcal{V} = 7.9 \times 10^7$ pc$^3$, and assuming a nominal value of the fraction $\eta = 0.1$ (e.g., Higdon et al. 1998; Tatischeff 2008) of the total SN energy, $E_\mathrm{ej} = 10^{51}$ erg per SN (Woosley & Weaver 1995), that is channeled to particle acceleration, the estimated CRp energy density is $u_p = \tau_\mathrm{res}\, \mathcal{V}^{-1}\, \mathcal{R}_\mathrm{SN}\, \eta\, E_\mathrm{ej}$. The CRp energy losses mainly occur through superwind advection and p–p interactions, with a characteristic residency timescale of $\tau_\mathrm{res} = (1/\tau_\mathrm{wind} + 1/\tau_\mathrm{pp})^{-1}$. In the latter expression, $\tau_\mathrm{wind} = r_s/v_\mathrm{wind} = 10^6 (r_\mathrm{s}/0.1\,\mathrm{kpc})(v_\mathrm{wind}/100\,\mathrm{km\,s^{-1}})^{-1}$ (Longair 1981), which, for $v_\mathrm{wind} = 500$ km s$^{-1}$ (Shopbell & Bland-Hawthorn 1998), gives $\tau_\mathrm{wind} = 6.8 \times 10^5$ yr. Also, $\tau_\mathrm{pp} = (\sigma_\mathrm{pp} c n_\mathrm{p})^{-1}$ which, for $\sigma_\mathrm{pp} \sim 40$ mb at a few tera-electronvolts (Kelner et al. 2006) and $n_\mathrm{p} = 200$ cm s$^{-1}$ (Table 5), is $\tau_\mathrm{pp} \sim 1.3 \times 10^5$ yr. Thus, $\tau_\mathrm{res} \sim 1.1 \times 10^5$ yr. The SN rates $\mathcal{R}_\mathrm{SN} \sim 0.13$ yr$^{-1}$ (consistent with X-ray and radio estimates, see Iwasawa 2021 and Lacki et al. 2011) yield

[7] The NT bremsstralung, computed from the thermal and NT electron densities, is largely subdominant even at its peak. This contradicts the corresponding result by the VERITAS Collab. et al. (2025). The discrepancy likely stems from the different CRe calibrations used in the two analyses.

$u_p \sim 385$ eV cm$^{-3}$, in accordance with our SED modeling results.

• *Radio spectrum.* Given that the CRp spectrum is fully determined by the measured $\gamma$-ray spectrum, and that consequently its secondary yield is also, the synchrotron radio spectrum mainly constrains the primary CRe spectrum. In the frequency range 0.3−9 GHz, the emission from the CSB region is fit by a thermally absorbed $\alpha_r = 0.62$ PL spectrum (Adebahr et al. 2013); thus, the emitting-CRe population has a spectral index of 2.24. The total emission is the superposed contributions by primary and secondary electrons in which the latter gets steeper than the total at $\nu \gtrsim 2$ GeV (Fig. 2). This feature motivates our assumption of $q_e \leq 2.2$ for primary CRe. Star formation is much less intense outside the CSR (Shopbell & Bland-Hawthorn 1998), so the influx of CR into this region is expected to be relatively week.

Owing to the varying lepton-to-hadron ratio (and therefore the primary-to-secondary electron ratio) as a function of the graybody model describing the FIR emission, the low-frequency thermal absorption, quantified by the emission measure (EM), also varies (Table 6). A quite minor $\nu^{-0.1}$ thermal free-free emission also enters the model spectrum, calibrated using the measured H$\alpha$ emission (Kennicutt et al. 2008) as a proxy.

• *CRe.* The deduced primary CRe spectrum, with PL index $q_e \lesssim 2.2$, may appear surprisingly flat. Actually, the fact that it is very close to the injection spectrum is understandable: irrespective of details of the acceleration mechanism, the electron injection timescale can be gauged by the inverse of the SN rate, $\tau_+ = \mathcal{R}_{\rm SN}^{-1}$ where $\mathcal{R}_{\rm SN} \sim 0.1$ yr$^{-1}$ (Iwasawa 2021). The CRe energy loss time is $\tau_- = \gamma/\dot{\gamma}$, where $\dot{\gamma}$ is the sum of loss rates by electronic excitations, bremsstrahlung, synchrotron, Compton (detailed, e.g., in Appendix A of Persic et al. 2024, and using the matter and radiation/magnetic-energy densities in the CSB reported in Tables 5, 6), and advection (Eq. 7). The comparison indicates that the injection timescale is much shorter than the loss time at all energies, even at $\gamma \sim \gamma_{\rm max}$. It is not surprising then that the primary electron component of the radio spectrum reflects the injection spectrum.

The production timescales of secondary CRe depend on the timescale of $\pi^\pm$ production following p−p interactions, $t_{\pi^\pm}$. The latter is $t_{\pi^\pm} = 1/(\kappa_{\pi^\pm} n_H \sigma_{\rm pp} c)$, where $\kappa_{\pi^\pm} = 0.25$ is the fraction of kinetic energy of the incident proton transferred to $\pi^\pm$; assuming $\sigma_{\rm pp} = 35$ mb as appropriate at a few tera-electronvolts (Kelner et al. 2006) and $n_{\rm H} = 200$ cm$^{-3}$ (Table 5), it is $t_{\pi^\pm} \simeq 1.9 \times 10^{13}$ s. This implies that the spectral distribution of secondary electrons mimics that of their parent CRp spectrum.

• *Magnetic field.* The value of the magnetic field ($B$) averaged over the CSB volume, i.e., co-spatial with the NT SED analyzed in this study, was derived by fitting the radio synchrotron emission generated by CRe whose spectrum was normalized to the NT 5 keV flux interpreted as Compton/FIR radiation ($j_{\rm Compt} \propto n_e$). This direct estimate of $B$ is the first of its kind for any CSB. The deduced $B$ values for our six models span the range $B = 97$–$143\,\mu$G (Table 6).

Our estimates of $B$ can be compared to earlier estimates that were typically based on the assumption of energy equipartition between CR and the magnetic field. As an example, in our model with $\beta = 1$, the magnetic field is $B = 97\,\mu$G and the CRp-to-CRe number density ratio is $p/e(1\,{\rm GeV}) = 120$. Interestingly, Adebahr et al. (2013) found a similar result based on the radio spectrum alone: assuming particle-field equipartition according to Beck & Krause's (2005) Eq. A18 with $p/e(1\,{\rm GeV}) = 120$, they obtained $B_{\rm eq} \approx 98\,\mu$G. Persic & Rephaeli (2014), too, estimated $B = 100\,\mu$G from radio emission, using a revised equipartition formula that includes theoretically based $p/e$ and $u_p/u_e$ ratios (including secondary CRe) as function of the CR spectral indices, as well as environment-dependent energy losses. Using their own revised equipartition formula that included pionic secondaries and energy losses, Lacki & Beck (2013) estimated $B = 240\,\mu$G. On the other hand, accounting for a single population of particles, Völk et al. (1989) estimated $B \sim 50\,\mu$G from minimum energy in CR plus the magnetic field. In their models of M82 multifrequency emission, Peretti et al. (2019) adopted $B = 210\,\mu$G, while De Cea del Pozo et al. (2009) found $B = 120 \div 290\,\mu$G accounting for uncertainties in the environmental parameters.

In all our models, the deduced values of $B$ are very close to its equipartition value, $B \approx B_{\rm eq} = \sqrt{8\pi\,(u_{\rm CRp} + u_{\rm CRe})}$, where $u_{\rm CRe}$ includes primary and secondary contributions. Thus, rather than assuming equipartition ab initio, we deduced its approximate validity in the CSB region.

• *Photon-photon absorption.* The intense local IR photon field raises the question of whether any photon–photon absorption affects the emerging $\gamma$-ray emission. The relevant cross section (Heitler 1960) is $\sigma_{\gamma\gamma}(E_\gamma, \epsilon, \phi) = (3/16)\,\sigma_{\rm T}\,(1-\beta^2)\,[2\beta(\beta^2-2) + (3-\beta^4)\ln[(1+\beta)/(1-\beta)]]$, where $\beta = \sqrt{1 - [2m^2c^4]/[E_\gamma \epsilon (1-\cos\phi)]}$, with $E_\gamma$ the energy of the incoming photon, $\epsilon$ the energy of the local photon, and $\phi$ the interaction angle between their trajectories. Integrating over $\phi$ and the spectrum of target photons, $n_{\rm IR}(\epsilon)$, and multiplying by the characteristic source size, $r_s$, we obtained the photon–photon optical depth, $\tau_{\gamma\gamma}(E_\gamma)$. Figure 3 shows that in situ photon–photon absorption is negligible for incoming energies of $\lesssim 10$ TeV, well above the energies sampled by the current VERITAS data.

• *Neutrino detectability.* It is of interest to assess the detectability of the $\pi^\pm$-decay neutrino emission from M82 with current and upcoming neutrino observatories. With an apparent dominant $\pi^0$-decay origin of $\gamma$ ray emission in the CSB, $\pi^\pm$-decay neutrinos are (obviously) also produced (e.g., Ambrosone et al. 2021). Our calculation of the predicted $\nu_\mu$ and $\nu_e$ spectra of M82, using Kelner et al.'s (2006) formalism (Sect. A.2.3 of Persic et al. 2024), indicates that the broadly peaked giga-electronvolt–tera-electronvolt neutrino flux (Fig. 4) is too low for detection by current and upcoming $\nu$ projects. This conclusion is based on the following calculation of the estimated observation time needed to detect M82 with an experiment with detection sensitivity comparable to the Antarctica-based IceCube+DeepCore Observatory, the most sensitive current $\nu$-detector at neutrino energies $E_\nu > 10$ GeV (e.g., Bartos et al. 2013). We also considered the Mediterranean-based KM3NeT observatory, set to become a sensitive detector of $E_\nu \sim 10$–$100$ GeV atmospheric neutrinos and of $E_\nu > 1$ TeV cosmic neutrinos (Adrián-Martinez et al. 2016).

► The IceCube+DeepCore $\mu_\nu$ effective area (Abbasi et al. 2012, Fig. 8-right) can be approximated as $A_{\rm eff}(E_{\nu_\mu}) = 10^{4-[0.65x^3-4.5x^2+6.95x+1.3]}$ cm$^2$, where $x = \log(E_{\nu_\mu})$ (energies in giga-electronvolts). ($A_{\rm eff}$ is ∼2 times smaller for $\nu_e$.) Only in the energy range 10 GeV–2.2 TeV do the Ice-Cube+DeepCore sensitivity and the M82 predicted diffuse spectral $\nu$-flux effectively overlap. In this band the $\nu$ flux can be approximated as $\frac{\mathrm{d}N_\nu}{\mathrm{d}E_\nu} \sim 10^{-8.34} E_{\nu_\mu}^{-q_p}$ cm$^{-2}$ s$^{-1}$ GeV$^{-1}$. The corresponding number of detected $\nu$ per year, $N_\nu = t_{1\,\mathrm{yr}} \int_{10\,\mathrm{GeV}}^{2.2\,\mathrm{TeV}} \frac{\mathrm{d}N_\nu}{\mathrm{d}E_\nu} A_{\mathrm{eff}}(E_{\nu_\mu})\, \mathrm{d}E_{\nu_\mu}$ (where 2.2 TeV is the cutoff of the neutrino spectrum, Fig. 4), is $N_{\nu_\mu} = 2.45\ \mathrm{yr}^{-1}$.

The detector background (for upward-going events) is dominated by atmospheric neutrinos produced by CR in the northern hemisphere. To compute such a noise background, $N_\nu$, we started from the atmospheric $\nu_\mu$ spectrum (Fig. 12 of Aartsen et al. 2015; the spectrum is $\sim$20 times lower), which in the energy range 10 GeV$-$2.2 TeV (see Fig. 4) can be approximated as $\phi(E_\nu) = E_\nu^{-2}\, 10^{-(ax^2+bx+c)}$ Gev$^{-1}$ s$^{-1}$ cm$^{-2}$ sr$^{-1}$ , where $a = -0.203457$, $b = 0.17507$, $c = 2.03243$, and $x = \log(E_\nu)$ with $E_\nu$ in giga-electronvolts. So the number of vertical neutrinos detected by IceCube+DeepCore is $N_\nu = t_{1\,\mathrm{yr}} \int_{10\,\mathrm{GeV}}^{2.2\,\mathrm{TeV}} \frac{\mathrm{d}N_\nu}{\mathrm{d}E_\nu} A_{\mathrm{eff}}(E_\nu)\, (\Delta\Omega_{\mathrm{M82}}/1.6\,\mathrm{sr})\, \mathrm{d}E_\nu \simeq 24\ \mathrm{yr}^{-1}$, for a solid angle, $\Delta\Omega_{\mathrm{M82}} \simeq 2.4 \times 10^{-4}$, corresponding to a 0.5 deg radius uncertainty in the direction reconstruction of IceCube+DeepCore $\nu_\mu$ toward M82. [8]

Based on this rough estimate, observations of diffuse giga-electronvolt–tera-electronvolt muon neutrinos from M82 would imply $S/N \sim 0.1$. This estimate indicates that the likelihood of detecting M82 with an IceCube+DeepCore–like detector is very low.

▶ The upcoming KM3NeT observatory is composed of two Cherenkov neutrino telescopes under construction in the Mediterranean sea, ARCA and ORCA (Astroparticle/Oscillation Research with Cosmics in the Abyss, respectively). ARCA, to be located 100 km southeast of Sicily's southernmost tip at 3500 m depth, is optimized for detection of $\gtrsim 1$ TeV astrophysical neutrinos. ORCA, to be located off the coast of Toulon (France) at a depth of 2500 m, is optimized for energies of $\approx$10 GeV in order to study neutrino properties exploiting neutrinos generated in the Earth atmosphere. The ARCA sensitivity for sources with an unbroken $E^{-2}$ spectrum for an observation time of 6 years is $\approx 10^{-9}$ GeV cm$^{-2}$ s$^{-1}$ for the full declination range $-1 \leq \sin(\delta) \lesssim 0.8$, similar to the results, for a similar exposure, reported by IceCube for northern hemisphere sources (Ajello et al. 2019). Clearly, ARCA would be the choice KM3NeT instrument to observe M82. Indeed, Ajello et al. (2024) have shown that individual galaxies with ongoing star formation, with concurrent AGN activity (NGC 1068 and the Circinus galaxy) or without (the Small Magellanic Cloud), can be detected by ARCA in 10 years of observation if their CRp spectra are hard ($q_p = 2$) and unbroken to very high energies (500 TeV). However, our modeling of the LAT and VERITAS $\gamma$-ray data for M82 implies $q_p \simeq 2.3$ and $E_p = 7$ TeV for the CRp spectrum, which translate into a $\approx$1 TeV cutoff in the M82 neutrino spectrum. At $E_\nu \approx 1$ TeV the ARCA effective area is as low as $A_{\mathrm{eff}} \approx 1\,\mathrm{m}^2$ (Adrián-Martinez et al. 2016), making ARCA unsuitable for detecting M82.

[8] The uncertainty is $\sim$5 deg at 100 GeV and somewhat greater at lower energies, but gets substantially smaller at higher energies: here we adopt 0.5 deg. The solid angle $1.6\pi$ corresponds to the solid angle in the northern hemisphere.

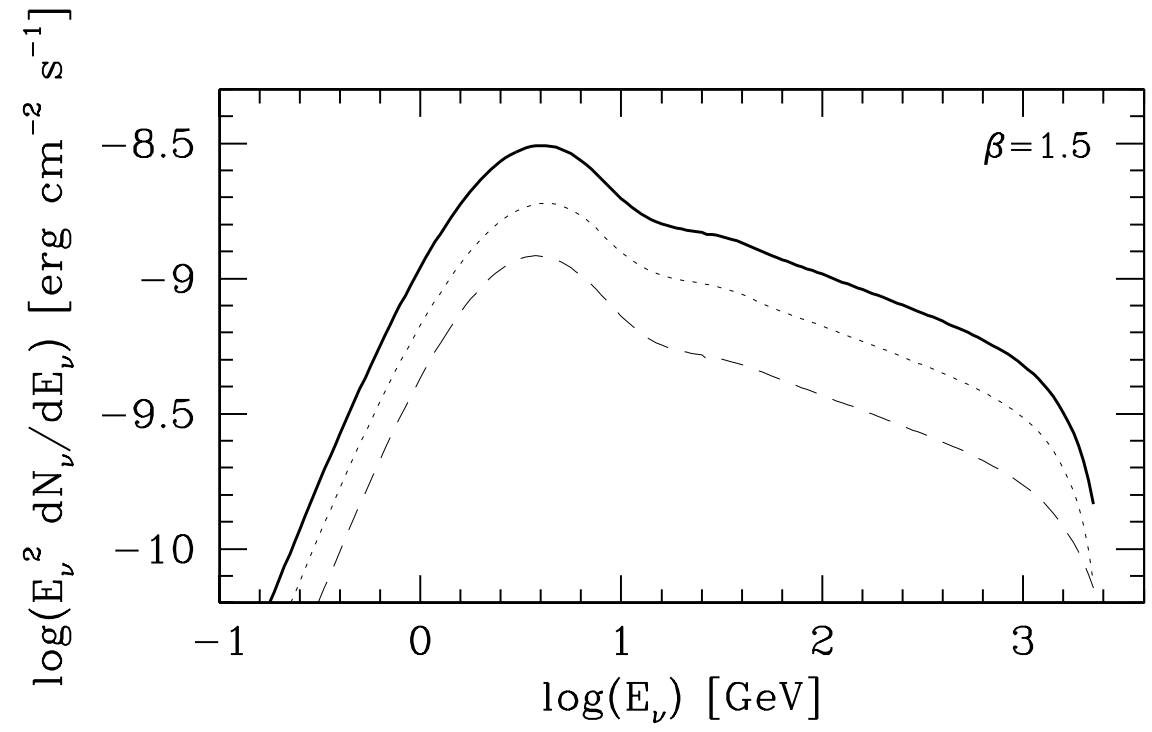

**Fig. 4.** Predicted neutrino spectral flux from M82 ($\beta = 1.5$ model). The neutrino spectra for all the models considered in this study look similar, due to the models' very similar hadronic characteristics. The curves represent total-$\nu$, $\nu_\mu$, $\nu_e$ spectra (thick solid, thin dotted, thin dashed lines, respectively).

## 6. Conclusion

The recent publication of diffuse NT 4$-$8 keV emission from M82's CSB motivates a different interpretation of the NT emission in this region. We complement the X-ray data with 50 MeV$-$16.25 TeV $\gamma$-ray data and 0.3$-$8.3 GHz radio data.

This analysis hinges on two points: first, identifying the high-energy $\gamma$-ray data as pionic emission: this fixes the CRp and secondary CRe spectra; and second, combining secondary and primary CRe to normalize the latter by fitting the predicted Compton/starlight emission to the NT X-ray flux and determine its spectral slope and cutoff from the low-energy $\gamma$-ray data. Observational uncertainties on the FIR graybody emission result in a range of viable SED models.

The SED models are quite similar in these models, the NT parameters only showing moderate variations. The CRp and primary-CRe spectra are PL with spectral indices of $q_p \approx 2.3$ and $q_e \approx 2.2$; the CR energy densities are $u_p \approx 385$ eV cm$^{-3}$ and $u_e \approx 5$ eV cm$^{-3}$, respectively. The magnetic field, deduced from a synchrotron fit to the radio spectrum, is $B \approx 120\,\mu$G, in energy equipartition with the CR. More precise measurements of the FIR emission would narrow down the range of viable NT SED models.

*Acknowledgements.* We thanks Dr Lab Saha of the VERITAS Collaboration for providing the tabulated 2008$-$2022 VERITAS spectral data (plotted in VERITAS Collaboration et al. 2025). MP acknowledges hospitality from the Physics & Astronomy Department of Padova University. We acknowledge useful and enlightening conversations with Jean Ballet and Philippe Bruel on some aspects of the LAT data analysis. The *Fermi*-LAT Collaboration acknowledges generous ongoing support from a number of agencies and institutes that have supportedboth the development and the operation of the LAT as well as scientific data analysis. These include NASA and the Department of Energy in the United States, the Commissariat à l'Énergie Atomique and the Centre National de la Recherche Scientifique/Institut National de Physique Nucléaire et the Physique des Particules en France, the Agenzia Spaziale Italiana (ASI) and the Istituto Nazionale di Fisica Nucleare (INFN) in Italy, the Ministry of Education, Culture, Sports, Science and Technology (MEXT) and the High-Energy Acceleration Research Organization (KEK) in Japan, and the K.A. Wallenberg Foundation and the Swedish National Space Board in Sweden. Additional support for science analysis during the operations phase is gratefully acknowledged from the Istituto Nazionale di Astrofisica (INAF) in Italy and the Centre National d'Études Spatiales in France.

## Appendix A: *Fermi*-LAT data analysis

The Large Area Telescope (LAT) is the main instrument on the *Fermi* Gamma-Ray Space Telescope. It comprises a silicon microstrip tracker, a cesium-iodide calorimeter, and a plastic anti-coincidence detector. The LAT covers an energy range from 20 MeV to ∼2 TeV with a field-of-view of 2.4 sr.

For the analysis described in this work we used 16.3 years of LAT data of the `P8R3_SOURCE` class. We considered events from 50 MeV to 300 GeV, within a region of interest centered on M82 and 15° in radius. We reduced the data set with the usual zenith angle cut < 90°, to limit the contamination from the bright Earth limb. The details of the LAT data selection are reported in Table A.1.

**Table A.1.** LAT data selection.

| | |
|---|---|
| Event class | `P8R3_SOURCE` |
| Time range | 2008/08/04 – 2024/11/24 |
| MET range | 239557417 – 754137291 |
| Energy range | 50 MeV – 300 GeV |
| ROI center | R.A., Dec.=148.94, 69.66 |
| ROI radius | 15° |
| Zenith angle | < 90° |

The global fit is performed with a binned likelihood analysis. The source model is based on the fourth LAT source catalog, data release 4, 4FGL-DR4 (Abdollahi et al. 2022; Ballet et al. 2023); the CSR cannot be resolved by the LAT and is modeled as a point source. Energy dispersion [9] and likelihood weighting (Bruel 2021) are taken into account in the analysis.

We started the analysis using a LogParabola spectrum for M82, following the spectral model used in the 4FGL. The spectral parameters $\alpha$ and $\beta$ we derived were compatible with 4FGL within uncertainties. If the analaysis is restricted to energies > 200 MeV the agreement is excellent. To investigate the presence of a rising spectrum at lower energies, located close to the expected pionic shoulder, we tried including a second spectral component with a PL spectrum (modelled as a second point source, coincident with M82). The additional component resulted very steep, with spectral index ∼5, and significant, with a TS= 29.5. The new component is non-negligible only at energies $\lesssim$ 200 MeV. Fixing the spectrum for M82 to the catalog values, test-statistics maps indicate that the excess is approximately point-like and centered on top of M82. The parameters for this component are listed in Table A.2.

There are several sources of systematic uncertainties to consider. Energy dispersion corrections were applied to the spectra. We proceeded by increasing the number of *edisp_bins* in the global fit to −3, after which the fit become insensitive to an additional increase. The uncommonly steep spectrum of the additional component at low energy magnifies any remaining uncertainty to an unprecedented level, though, and the usual estimates on the residual systematic uncertainties may be too optimistic. M82 is located close to the north celestial pole. In proximity of the celestial poles, there is an increased risk of contamination from photons from the Earth limb, especially at low energies due to the broader PSF. A loop of interstellar gas, part of the North Celestial Pole Loop, passes very close to M82 and a relatively brighter globular structure, much smaller than the LAT point spread function at low energies, lies very close to M82. It is worth noting that the gas structure in question may be a nearby SNR (Marchal & Martin 2023; Schmelz et al. 2023). Uncertainties in the emission model can affect the spectral estimate for M82 at low energy; uncertainties in modelling the diffuse models were addressed in part by applying log-likelihood weights.

To try and address the aforementioned issues we performed an additional analysis of the low-energy part of the spectrum, selecting only events belonging to the PSF3 class: LAT events are subdivided into quartiles depending on the quality of the direction reconstruction, and we restrict ourself to those with the best PSF. We divided the events in the energy range 50–213 MeV in four logarithmic bins, and proceeded with the binned likelihood analysis. In the first two bins, M82 is now not detectable with any significance, but the upper limits we derive are still relatively high. The difference with the previous estimates indicate severe contamination from sources other than M82. In bin 3, M82 is detected with TS=20, and an excess with respect to the 4FGL spectrum is significant with TS=6. In bin 4, M82 is detected with TS=17 and the derived spectrum is compatible within uncertainties with the 4FGL value.

In this work we consider the 4FGL spectrum above 213 MeV. Below that energy, we have indications of an additional component, but systematics make it impossible to derive a spectral shape. We consider the 4FGL spectrum as a lower bound, and our best estimate (PSF3 U.L.) as an upper bound.

The SED was evaluated by dividing the whole energy range in 12 equally spaced logarithmic energy bins, and the first two low-energy bins were further divided in two, and finally by performing a dedicated binned likelihood fit in each bin. In the broader bins, we used the broader selection (`SOURCE` event class). The spectra of M82 was described as a PL with the fixed spectral index corresponding to the 4FGL spectrum considering the bin energy centroid. The SED data points resulted perfectly compatible with the ones distributed with 4FGL and with previous analyses (Ajello et al. 2020). In the finer, low-energy bins, we used the restrictive selection (`SOURCE-PSF3` event class). We evaluated upper limits with the standard binned likelihood approach. The results are shown in Fig. A.1. The dashed line represents the 4FGL spectrum, the filled gray band indicates the 1-$\sigma$ uncertainty. The data points are the SED we derived in this work, assuming the spectral shape above but using 16.3 years of data; the error bars indicate 1-$\sigma$ uncertainties. The upper bounds are the PSF3 90% C.L. upper limits. The lower bounds at low energy are the 4FGL spectral values, evaluated at the bin energy centroids. The resulting LAT flux values and upper/lower limits are reported in Table A.3.

**Table A.2.** Spectrum parameters for the additional low-energy PL component, as evaluated for the standard *SOURCE* selection.

| | |
|---|---|
| Prefactor $[10^{-9}\ \mathrm{ph/cm^2/s}]$ | $7.8 \pm 1.5$ |
| Index | $5.25 \pm 0.61$ |
| $E_{scale} = E_{decorr}$ | 50.82 |
| TS | 29.5 |

[9] https://fermi.gsfc.nasa.gov/ssc/data/analysis/documentation/Pass8_edisp_usage.html

**Table A.3.** *Fermi*-LAT spectral points. Data points: SED points assuming 4FGL spectrum. U.L.: 90% C.L. upper limits from the PSF3 analysis. L.L.: 4FGL fluxes at the same energies as the U.L.'s.

| Energy MeV | Flux ± Error $10^{-12}$erg cm$^{-2}$s$^{-1}$ | L.L. $10^{-12}$erg cm$^{-2}$s$^{-1}$ | U.L. |
|---|---|---|---|
| 59.93 | | >1.48 | <6.29 |
| 86.12 | | >1.63 | <4.55 |
| 123.74 | | >1.77 | <3.99 |
| 177.81 | | >1.88 | <2.54 |
| 306.26 | $1.99 \pm 0.19$ | | |
| 632.31 | $1.61 \pm 0.13$ | | |
| 1305.50 | $1.78 \pm 0.12$ | | |
| 2695.40 | $1.54 \pm 0.13$ | | |
| 5565.04 | $1.02 \pm 0.14$ | | |
| 11489.85 | $0.94 \pm 0.19$ | | |
| 23722.50 | $0.70 \pm 0.23$ | | |
| 101123.56 | $0.69 \pm 0.44$ | | |
| 208784.52 | $0.53 \pm 0.40$ | | |

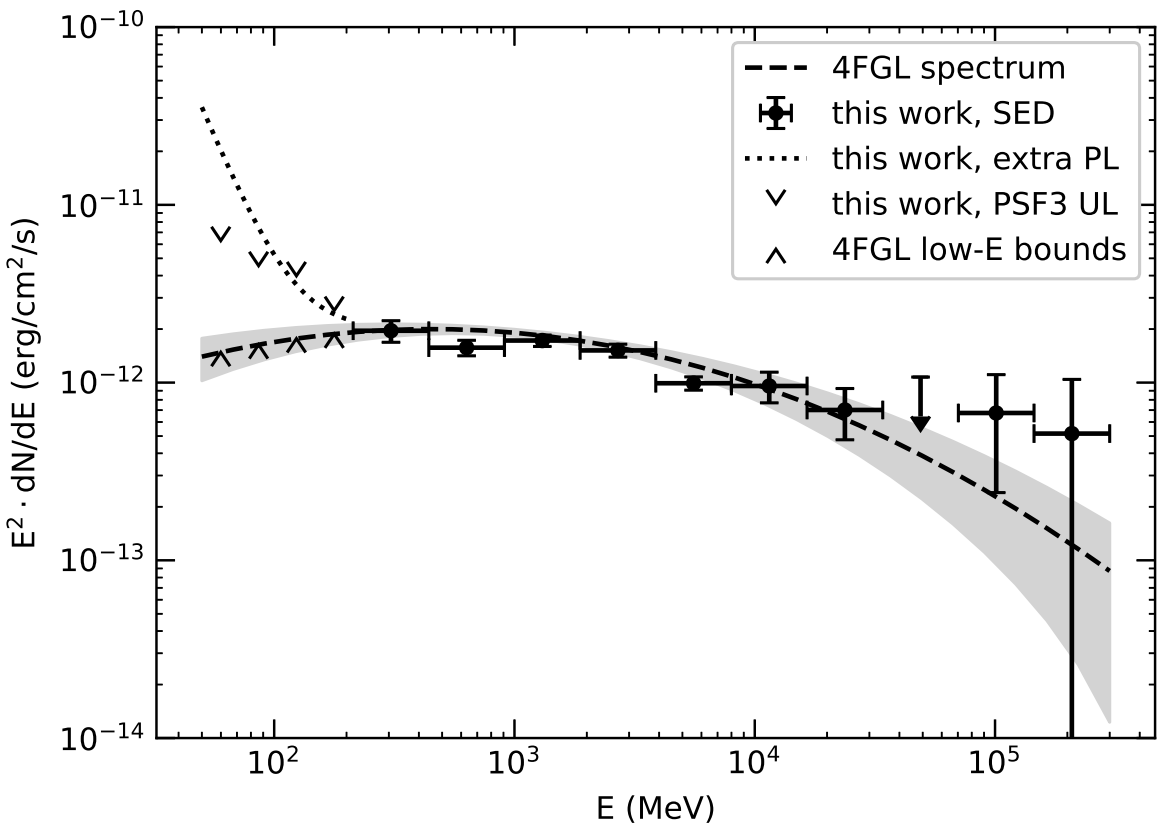

**Fig. A.1.** M82 spectral model. Dashed line: 4FGL spectrum with $1\sigma$ uncertainties (gray band). Data points: SED points derived assuming 4FGL spectrum, with $1\sigma$ uncertainties. Dotted line: 4FGL spectrum plus the low-energy PL component. Downward triangles: low-energy upper limits from the PSF3 analysis. Upward triangles: low-energy lower bounds, flux values from 4FGL model.

## Appendix B: VERITAS data

VERITAS is a stereoscopic system of four 12-m–diameter imaging atmospheric Cherenkov telescopes, located at Whipple Observatory in southern Arizona, USA at an elevation of 1268 m above sea level. Its energy threshold is ∼100 GeV, and its sensitivity (i.e., $5\sigma$ detection) is ∼0.6% of the Crab Nebula flux in 50 hours of observation at small (<30°) zenith angles (Park & VERITAS Collaboration 2015). Table B.1 lists the VERITAS spectral points described and plotted in VERITAS Collaboration et al. (2025; see their Fig. 1).

**Table B.1.** 2008−2022 VERITAS data.

| $E_{min}$ TeV | $E_{max}$ TeV | E TeV | F ± dF [cm$^2$s TeV]$^{-1}$ | U.L. [cm$^2$s TeV]$^{-1}$ |
|---|---|---|---|---|
| 0.3500 | 0.5137 | 0.4240 | 1.30E-12 ± 9.61E-13 | |
| 0.5137 | 0.7541 | 0.6224 | 3.72E-13 ± 2.10E-13 | |
| 0.7541 | 1.1068 | 0.9136 | 1.90E-13 ± 7.10E-14 | |
| 1.1068 | 1.6246 | 1.3409 | 8.87E-14 ± 2.99E-14 | |
| 1.6246 | 2.3845 | 1.9682 | 4.34E-14 ± 1.39E-14 | |
| 2.3845 | 3.5000 | 2.8889 | 9.12E-15 ± 6.36E-15 | |
| 3.5000 | 5.1373 | 4.2403 | 5.23E-15 ± 3.20E-15 | |
| 5.1373 | 7.5405 | 6.2240 | | <4.48E-15 |
| 7.5405 | 11.068 | 9.1355 | | <1.68E-15 |
| 11.068 | 16.2456 | 13.4092 | | <6.74E-16 |